\documentclass[twocolumn]{jpsj3}
\usepackage{color}

\title{Strong-Coupling Theory of Rattling-Induced Superconductivity}

\author{Kunihiro {\sc Oshiba}$^{1}$ and Takashi {\sc Hotta}$^{1,2}$}

\inst{$^{1}$Department of Physics, Tokyo Metropolitan University,
Hachioji, Tokyo 192-0397, Japan \\
$^{2}$Advanced Science Research Center, Japan Atomic Energy Agency,
Tokai, Ibaraki 319-1195, Japan}

\recdate{\today}

\abst{
In order to clarify the mechanism of the enhancement of
superconducting transition temperature $T_{\rm c}$
due to anharmonic local oscillation of a guest ion
in a cage composed of host atoms, i.e., {\it rattling},
we analyze the anharmonic Holstein model
by applying the Migdal-Eliashberg theory.
From the evaluation of the normal-state electron-phonon
coupling constant,
it is found that the strong coupling state is developed,
when the bottom of a potential for the guest ion becomes
wide and flat.
Then, $T_{\rm c}$ is enhanced with the increase of
the anharmonicity in the potential,
although $T_{\rm c}$ is rather decreased
when the potential becomes a double-well type
due to very strong anharmonicity.
From these results, we propose a scenario of anharmonicity-controlled
strong-coupling tendency for superconductivity induced by rattling.
We briefly discuss possible relevance of the present scenario
with superconductivity in $\beta$-pyrochlore oxides.
}

\kword{rattling, superconductivity, heavy electron, cage materials,
$\beta$-pyrochlore}

\begin{document}
\maketitle

\section{Introduction}

Recently, materials with cage structure have attracted much attention
in the research field of condensed matter physics from the emergence of
exotic magnetism and novel superconductivity.
The materials are, for instance,
filled skutterudites,\cite{Review,Goto,Kotegawa}
$\beta$-pyrochlore oxides,\cite{Yonezawa1,Yonezawa2,Yonezawa3,Hiroi1}
and clathrate compounds.\cite{Avila1,Avila2,Sales}
A common feature of this material group is the existence of nano-size cage
composed of relatively light atoms.
The ion contained in the cage, frequently called a guest ion,
feels a highly anharmonic potential and it can oscillate
with large amplitude.
Such oscillation of the guest ion in the cage is called
{\it rattling}, which is expected to be an origin of
interesting physical properties of cage materials
such as magnetically robust heavy electron state
of filled skutterudites,\cite{SmOsSb}
superconductivity of $\beta$-pyrochlore oxides,\cite{Hiroi2}
and high thermoelectric performance of clathrate compounds.\cite{Saramat}

From a theoretical viewpoint, heavy electron state with
phonon origin has been actively investigated in recent years.
The magnetically robust heavy electron phenomenon has been discussed
from the viewpoint of the Kondo effect with phonon origin.
\cite{Miyake,Hattori1,Hattori2}
The periodic Anderson-Holstein model has been analyzed
with the use of a dynamical mean-field theory and
a mechanism of the mass enhancement due to large lattice fluctuations
and phonon softening towards double-well potential
has been addressed.\cite{Mitsumoto}
One of the present authors has performed numerical calculations
for the Anderson model coupled with local Jahn-Teller and
Holstein phonons.\cite{Hotta0a,Hotta0b,Hotta0c,Hotta0d}
From the evaluation of electronic specific heat coefficient
$\gamma_{\rm e}$ in the Anderson-Holstein model,
it has been found that $\gamma_{\rm e}$ is enhanced by
the rattling, namely, the oscillation of the guest ion
in a potential with wide and flat bottom.
Then, it has been shown that $\gamma_{\rm e}$ enhanced by the rattling
is actually robust for an applied magnetic field.\cite{Hotta1}
Furthermore, it has been pointed out that the Kondo effect
due to the rattling exhibits peculiar isotope effect and
it can be the experimental evidence for rattling-induced
heavy fermion phenomenon.\cite{Hotta2}

Now we turn our attention to rattling-induced phenomena
in $\beta$-pyrochlore oxides AOs$_{2}$O$_{6}$ (A=K, Rb, and Cs).
It has been observed that $T_{\rm c}$ increases with the decrease
of radius of A ion: $T_{\rm c}$=9.6K for A=K,
$T_{\rm c}$=6.4K for A=Rb,
and $T_{\rm c}$=3.25K for A=Cs.\cite{Hiroi2}
The origin of this difference in $T_{\rm c}$ has been expected
to be the anharmonic oscillation of A ion.
In fact, it has been revealed that the anharmonicity of the potential
for A ion is enhanced when we change A ion in the order of
Cs, Rb, and K.\cite{Kunes}
In particular, KOs$_{2}$O$_{6}$ has relatively high $T_{\rm c}$
in comparison with other $\beta$-pyrochlore oxides
and it exhibits a remarkable first order transition
without symmetry change at $T_{\rm p} = 7.5K$.\cite{Hiroi3}
In $\beta$-pyrochlore oxides, characteristic behavior
has been observed in the temperature dependence of
electrical resistivity \cite{Hiroi1a}
and NMR relaxation rate.\cite{Yoshida}

These interesting phenomena in $\beta$-pyrochlore oxides
have been considered to originate from the rattling, i.e.,
anharmonic oscillation of alkali ion in the cage.
Dahm and Ueda have explained unusual temperature dependence of
electrical resistivity and NMR relaxation rate
by anharmonic phonons.\cite{Dahm}
Hattori and Tsunetsugu have investigated more realistic model
including three dimensional anharmonic phonons
in tetrahedral symmetry and have found that $T_{\rm c}$ is
strongly enhanced with increasing the third-order anharmonicity
of the potential.\cite{Hattori0,Hattori}
Chang {\it et al.} have discussed the superconductivity and
the first-order transition in KOs$_{2}$O$_{6}$
by using the strong-coupling Eliashberg approach in
the anharmonic Hamiltonian including forth-order terms.\cite{Chang}
As for the first-order transition observed in KOs$_{2}$O$_{6}$,
Fuse and \=Ono have proposed a possible scenario
on the basis of bipolaron formation.\cite{Fuse}
From these efforts, it has been gradually recognized that
the anharmonic oscillation of the ion in the cage is
closely related to the emergence of superconductivity
with relatively high $T_{\rm c}$.
However, the mechanism of the enhancement of $T_{\rm c}$
due to anharmonicity is still unclear.
In particular, it is worth while investigating the relation
among $T_{\rm c}$, mass enhancement, and anharmonicity
in the same model Hamiltonian.

In this paper, we analyze the anharmonic Holstein model
in an adiabatic approximation,
in order to understand the enhancement of
superconducting transition temperature $T_{\rm c}$
and the electron mass enhancement.
For the purpose, we calculate the electron self-energy
by using standard Green's function method and evaluate $T_{\rm c}$
by solving the Eliashberg equation within the Migdal approximation.
\cite{Migdal, Eliashberg}
Due to the evaluation of an effective electron-phonon coupling
constant $\lambda_{\rm n}$, it is found that $\lambda_{\rm n}$
is increased with the increase of the anharmonicity even if the original
electron-phonon coupling is weak.
This strong-coupling tendency due to anharmonicity induces
the enhancement of electron effective mass and $T_{\rm c}$.
Then, it is shown that $T_{\rm c}$ becomes maximum
when the guest ion oscillates with large amplitude
in a potential with a wide and flat bottom.
We also find that the magnitude of $T_{\rm c}$ is increased
with the increase of the Debye-Waller factor,
as observed experimentally in $\beta$-pyrochlore oxides.

The organization of this paper is as follows.
In Sec.~2, we introduce the model Hamiltonian and
discuss briefly the relation between potential shape and anharmonicity.
Then, we provide the formulation to evaluate electron self-energy
and $T_{\rm c}$ within a framework of the Migdal-Eliashberg theory.
In Sec.~3, we show our calculated results and discuss the key role of
anharmonicity for the emergence of rattling-induced superconductivity.
We also discuss possible relevance of our scenario with
experimental results of $\beta$-pyrochlore oxides.
Finally, we summarize this paper in Sec.~4.
Throughout this paper, we use such units as $\hbar$=$k_{\rm B}$=1.

\section{Model and Formulation}

\subsection{Anharmonic Holstein Model}

Let us introduce the anharmonic Holstein model in which
conduction electrons are coupled with local oscillation of guest ion
in the anharmonic potential.
The Hamiltonian consists of three parts as
\begin{equation}
   \label{hamiltonian}
   H = \sum_{\mib{k},\sigma} \varepsilon_{\mib{k}}
       c_{\mib{k}\sigma}^{\dag} c_{\mib{k}\sigma}
     + H_{\rm eph} + H_{\rm ph},
\end{equation}
where $\varepsilon_{\mib{k}}$ is the energy of conduction electron,
$\mib{k}$ is momentum, $c_{\mib{k} \sigma}$ indicates an annihilation operator of
electron with spin $\sigma$ and momentum $\mib{k}$,
$H_{\rm eph}$ indicates the electron-phonon coupling,
and $H_{\rm ph}$ is the phonon part.
Throughout this paper, we consider a half-filling situation
by appropriately adjusting the value of a chemical potential
included in $\varepsilon_{\mib{k}}$,
although we do not explicitly include the chemical potential term
in $H$.

The second term $H_{\rm eph}$ in eq.~(\ref{hamiltonian}) indicates
the electron-phonon coupling part, given by
\begin{equation}
  H_{\rm eph} = g \sum_{{\bf i},\sigma}
  c_{{\bf i}\sigma}^{\dag} c_{{\bf i}\sigma} Q_{\bf i},
\end{equation}
where $g$ denotes an electron-phonon coupling constant,
${\bf i}$ indicates an ion site,
$c_{{\bf i} \sigma}$ is the annihilation operator of electron
at a site ${\bf i}$,
and $Q_{\bf i}$ is normal coordinate of oscillation of
the guest ion at a site ${\bf i}$.
Note here that the reduced mass for the normal oscillation
is set as unity in this paper for simplicity.

The third term $H_{\rm ph}$ in eq.~(\ref{hamiltonian})
denotes the phonon part, written as
\begin{equation}
  H_{\rm ph} = \sum_{\bf i} [P^2_{\bf i}/2+V(Q_{\bf i})],
\end{equation}
where $P_{\bf i}$ denotes the canonical momentum of the guest
ion at a site ${\bf i}$
and $V(Q_{\bf i})$ indicates the potential for the guest ion.
In order to include the effect of anharmonicity,
we express $V(Q_{\bf i})$ as
\begin{eqnarray}
  V(Q_{\bf i}) = \omega_{0}^2 Q_{\bf i}^{2}/2
  +k_{4} Q_{\bf i}^{4}+k_{6} Q_{\bf i}^{6},
\end{eqnarray}
where $\omega_{0}$ is the energy of the guest ion
and $k_{4}$ and $k_{6}$ denote the coefficients for
fourth- and sixth-order anharmonic terms, respectively.

For quantum statistical calculations,
it is convenient to introduce the annihilation operator of phonon
through the standard procedure as
$Q_{\bf i} = (a_{\bf i} + a_{\bf i}^{\dagger})/ \sqrt{(2\omega_{0})}$.
Then, we obtain $H_{\rm eph}$ and $H_{\rm ph}$, respectively, as
\begin{equation}
  H_{\rm eph} = \sqrt{\alpha} \omega_{0} \sum_{{\bf i}, \sigma}
  (a_{\bf i}^{\dagger}+a_{\bf i})c_{{\bf i}\sigma}^{\dag} c_{{\bf i}\sigma},
\end{equation}
and
\begin{equation}
  \label{hamiltonian-phonon-part-field-rep}
  H_{\rm ph} \!=\! \omega_{0} \sum_{\bf i} [a_{\bf i}^{\dagger}a_{\bf i}
  \!+\! 1/2
  \!+\! \beta(a_{\bf i}+a_{\bf i}^{\dagger})^4
  \!+\! \gamma(a_{\bf i}+a_{\bf i}^{\dagger})^{6}],
\end{equation}
where $\alpha$, $\beta$, and $\gamma$ are given by
\begin{equation}
 \alpha = \frac{g^{2}}{2\omega_{0}^{3}},~
  \beta = \frac{k_{4}}{4\omega_{0}^{3}},~
 \gamma = \frac{k_{6}}{8\omega_{0}^{4}},
\end{equation}
respectively.
Note that $\alpha$ is non-dimensional electron-phonon coupling
constant, while $\beta$ and $\gamma$ indicate non-dimensional
fourth- and sixth-order anharmonicity parameters, respectively.
By using these non-dimensional parameters,
we express the potential as
\begin{eqnarray}
  V(q_{\bf i}) = \alpha \omega_{0}(q_{\bf i}^{2}
  + 16 \alpha \beta q_{\bf i}^{4}
  + 64 \alpha^{2} \gamma q_{\bf i}^{6}),
\end{eqnarray}
where $q_{\bf i} = Q_{\bf i}/\ell$
and $\ell$ denotes a typical length scale of the oscillation,
given by $\ell = g/\omega_{0}^{2}$.

\begin{figure}[t]
\begin{center}
\includegraphics[width = 85mm , angle = 0]{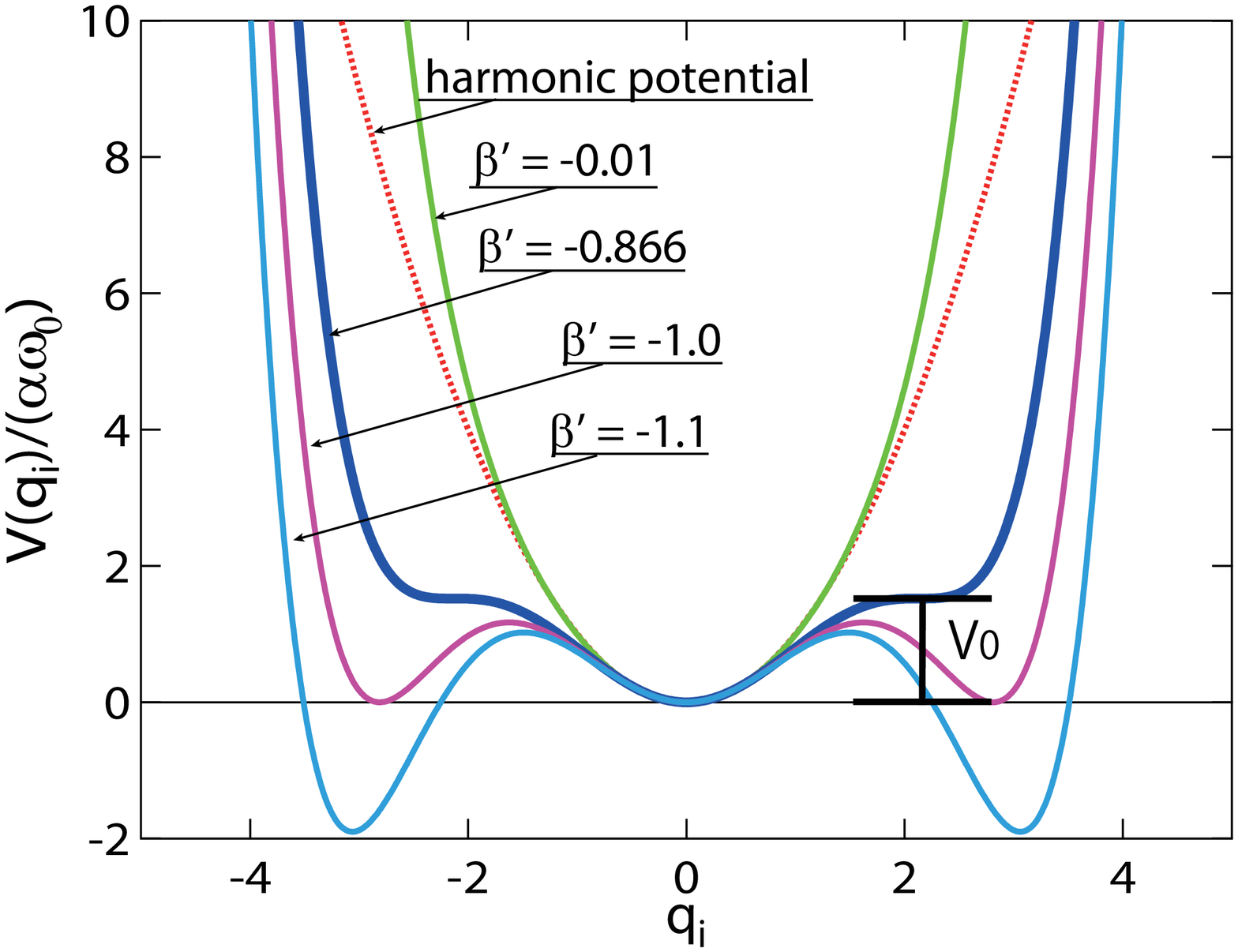} 
\end{center}
\caption{(Color online) Anharmonic potentials for $\gamma=10^{-5}$.
We show the harmonic potential as a dotted curve for comparison.
The curve of $\beta'=-0.01$ denotes the on-center type,
while that of $\beta'=-1.2$ indicates the off-center one.
As for the meaning of $V_0$, see the main text.
}
\label{fig1}
\end{figure}

Before proceeding to the formulation,
here we briefly discuss how the potential shape is changed by
anharmonicity parameters.
It is noted that we consider only the case of $\gamma > 0$,
since the oscillation of the guest ion should be confined
in a finite space of the cage.
In order to characterize the potential shape,
we define the re-scaled anharmonicity parameter as
\begin{equation}
   \beta' = \frac{\beta}{\sqrt \gamma}.
\end{equation}
With the use of this parameter,
even if the value of $\gamma$ is changed,
the potential shape can be discussed in the same range of $\beta'$.
As already mentioned in previous papers,\cite{Hotta1,Hotta2,oshiba}
the potential shapes are classified into three types by
the values of $\beta'$ as
on-center type for $\beta' > -\sqrt{3}/2$,
rattling type for $-\sqrt{3}/2 \geq \beta' \geq -1$,
and off-center type for $\beta' < -1$.
Typical potential shapes are shown in Fig.~\ref{fig1}
for various values of $\beta'$.
Note that the potential height $V_{0}$ is defined as
the difference between the potential values at local maxima
and minima.
As a typical example, we show $V_{0}$ for the case of
$\beta'=-\sqrt{3}/2$.
Here we note that $\beta'=-\sqrt{3}/2$ is defined by the value
at which the potential exhibits saddle points
at $q_{\bf i} \approx \pm 2$,
while $\beta'=-1$ indicates the case for which
all the potential values at three minima become zeros.
It is observed that the rattling-type potential has relatively
wide and flat region in the bottom in comparison with other two cases.
Namely, in this case, the guest ion is considered to oscillate
with large amplitude in the anharmonic potential.
From the purpose of this paper, it is quite natural to focus on
the parameters around the values for the rattling-type potential.

\subsection{Formulation}

In order to discuss the superconducting transition temperature $T_{\rm c}$,
we analyze the Hamiltonian by applying the formalism of
the Migdal-Eliashberg theory.\cite{Migdal,Eliashberg}
First we consider the normal self-energy in the Migdal approximation
in which electron-phonon vertex corrections are ignored in an adiabatic limit
of $\omega_0 \ll W$,
where $\omega_0$ denotes the phonon energy and $W$ is the electron bandwidth.
Then, in the second-order perturbation theory in terms of $g$,
the self-energy $\Sigma$ is given by
\begin{eqnarray}
  \label{electron_self_energy_anharmonic}
  \Sigma(i\omega_{n}) = -\alpha \omega_{0}^{2} T\sum_{n'}
  \sum_{\mib{k'}} D_{0}(i\omega_{n}-i\omega_{n'})
  G(\mib{k'},i\omega_{n'}),
\end{eqnarray}
where $T$ is a temperature,
$\omega_{n}$ is the fermion Matsubara frequency defined by
$\omega_{n} = (2n+1)\pi T$ with an integer $n$,
$D_{0}$ is the phonon Green's function shown below,
and $G(\mib{k},i\omega_{n})$ is the electron Green's function,
given by
\begin{eqnarray}
  G(\mib{k},i\omega_{n}) =
  \frac{1}{i\omega_{n}-\varepsilon_{\mib{k}}-\Sigma(i\omega_{n})}.
\label{electron's_Green's_function}
\end{eqnarray}
We note that the site dependence dose not appear in $D_{0}$,
since we consider Einstein-type local phonons in
the adiabatic approximation.
Thus, the self-energy does not depend on the momentum.
Note also that in our calculation, we use the bare phonon Green's function
$D_{0}$, instead of dressed phonon Green's function $D$,
since we consider an adiabatic situation in which
$W$ is much larger than the phonon energy $\omega_{0}$.

Concerning the phonon Green's function $D_0$ of anharmonic system,
it is convenient to express it in the spectral representation as
\begin{equation}
 D_{0}(i\nu_{n}) = \int \frac{\rho_{\rm ph}(\omega)}{i\nu_{n}-\omega} d\omega,
\label{phonon_green}
\end{equation}
where $\nu_{n}$ is the boson Matsubara frequency defined
by $\nu_{n} = 2n\pi T$,
and $\rho_{\rm ph}(\omega)$ is phonon spectral function,
given by
\begin{equation}
 \rho_{\rm ph}(\omega) = \sum_{i,j} A_{i,j} \delta(\omega+E_i-E_j).
\end{equation}
Here $E_i$ is the $i$-th eigenenergy of $H_{\rm ph}$
and the spectral weight $A_{i,j}$ is given by
\begin{equation}
 A_{i,j}= \frac{1}{\Omega}
  (e^{- E_{i}/T}-e^{- E_{j}/T})
  | \langle \Phi_{i} |u_{\bf i}|\Phi_{j} \rangle |^{2},
\end{equation}
where $|\Phi_{i} \rangle$ is the $i$-th eigenstate of $H_{\rm ph}$,
$\Omega$ is the partition function, given by
$\Omega=\sum_i e^{-E_{i}/T}$,
and $u_{\bf i}$ is the non-dimensional displacement operator,
given by $u_{\bf i}$=$a_{\bf i}+a_{\bf i}^{\dagger}$.

Let us define an electron mass enhancement factor $Z$,
given by
\begin{eqnarray}
  Z = 1-\frac{\Sigma (i\pi T)}{i \pi T}.
\end{eqnarray}
We note that the effective mass $m^*$ is given by $m^{*}$=$Zm$,
where $m$ is bare electron mass.
Here we assume that $Z$ is given by the difference of $\Sigma$
on the imaginary axis at the lowest temperature,
although $Z$ should be defined by the energy
differentiation of $\Sigma$ on the real axis as
\begin{eqnarray}
  Z = 1-\frac{\partial {\rm Re} \Sigma(\omega)}{\partial \omega}
        \bigg|_{\omega=0},
\end{eqnarray}
where $\Sigma(\omega)$ denotes the electron self-energy on the real axis
and $\omega$ denotes the energy.
Since $\Sigma(\omega)$ is found to exhibit a peak in the order of the
phonon energy $\omega_{0}$, we replace the energy differentiation
with the energy difference of the width $T$ for $T < \omega_{0}.$\cite{oshiba}
In this paper, we are interested in the quantity around $T=T_{\rm c}$
and $T_{\rm c}$ is considered to be lower than $\omega_{0}$.
Thus, we adopt the present definition for $Z$.

For the purpose to evaluate $T_{\rm c}$, we solve the linearized
gap equation at $T$=$T_{\rm c}$, given by
\begin{eqnarray}
  \phi(i\omega_{n}) = \alpha \omega_{0}^{2} T\sum_{n'} \sum_{\mib{k}'}
  D_{0}(i\omega_{n}-i\omega_{n'}) F(\mib{k}',i\omega_{n'}),
  \label{anomalous_self_energy_anharmonic}
\end{eqnarray}
where $\phi(i\omega_{n})$ is anomalous self-energy
and $F(\mib{k},i\omega_{n})$ is anomalous Green's function,
given in the linearized form in the vicinity of $T_{\rm c}$ as
\begin{eqnarray}
  \label{F}
  F(\mib{k}, i\omega_{n}) = -G(\mib{k},i\omega_{n})
  G(-\mib{k},-i\omega_{n})\phi(i\omega_{n}).
\end{eqnarray}
Note that here we assume the isotropic $s$-wave gap,
since we consider Cooper pairs mediated by phonons.
Thus, the momentum dependence is suppressed
in the anomalous self-energy.

The calculation procedure is as follows:
First we determine the normal self-energy by solving
self-consistently eqs.~(\ref{electron_self_energy_anharmonic})
and (\ref{electron's_Green's_function}).
Then, by using the obtained normal self-energy,
we solve the gap equation
eqs.~(\ref{anomalous_self_energy_anharmonic}) and (\ref{F}).
Note that $T_{\rm c}$ is determined as a temperature
at which the positive maximum eigenvalue of
the gap equation becomes unity.

In the actual calculations,
we assume the electron density of states with
rectangular shape of the electron bandwidth $W$.
Hereafter $W$ is taken as the energy unit and we set $W=1$.
For the sum on the imaginary axis,
we use $32768$ Matsubara frequencies.
In order to accelerate the actual calculations in
eqs.~(\ref{electron_self_energy_anharmonic})
and (\ref{anomalous_self_energy_anharmonic}),
we exploit a Fast-Fourier-Transformation algorithm
for the summation in terms of $n'$.
The eigenvalue of the gap equation is obtained with
the use of the power method.
For diagonalization of $H_{\rm ph}$, we use 250 phonon basis.
Note that we have checked the convergence of the eigenenergy
by comparing the results obtained by using 300 phonon basis.
Throughout this paper, we set $\omega_{0} = 0.1$ and
$\lambda = 0.5$, where $\lambda$ is the Eliashberg electron-phonon
coupling constant given by $\lambda = 2 \alpha \omega_{0} / W$.

\section{Calculated Results}

\subsection{Superconducting transition temperature}

\begin{figure}
\begin{center}
\includegraphics[width=85mm,angle = 0]{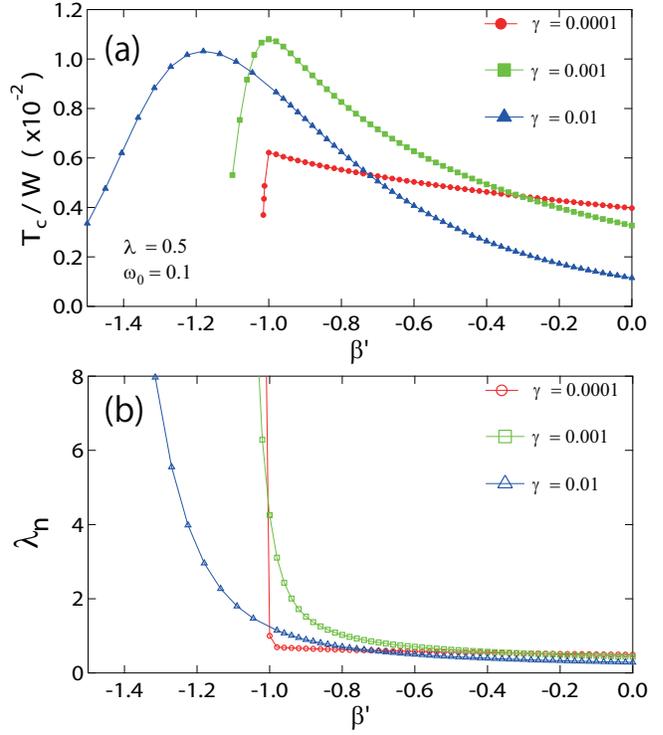}
\end{center}
\caption{(Color online)
(a) Superconducting transition temperature $T_{\rm c}$ vs. $\beta'$
for $\gamma$=$0.01$, $0.001$, and $0.0001$.
Here we set $\lambda = 0.5$ and  $\omega_{0} = 0.1$.
(b) Effective electron-phonon coupling constant $\lambda_{\rm n}$
vs. $\beta'$ for the same parameters as in the panel (a).
}
\label{fig2}
\end{figure}

Let us first discuss the effect of anharmonicity on
the superconducting transition temperature $T_{\rm c}$.
In Fig. \ref{fig2}(a), we depict three curves of $T_{\rm c}$
vs. $\beta'$ for $\gamma$=$0.01$, $0.001$, and $0.0001$.
For $\gamma$=$0.0001$, $T_{\rm c}$ slowly increases
with the decrease of $\beta'$
in the range of $\beta'>-1$ and it rapidly decreases at
$\beta'=-1$.
Note that we could not calculate $T_{\rm c}$ smaller than $0.001$
due to the limitation of the present numerical calculations.
For $\gamma=0.001$, we also observe that $T_{\rm c}$ is enhanced
by the increase of fourth-order anharmonicity and it takes
the largest value at $\beta' \simeq -1$.
Note that the largest value is also the highest among those
in three curves.
In the parameter region of the off-center type potential,
$T_{\rm c}$ rapidly decreases.

For $\gamma = 0.01$, $T_{\rm c}$ also exhibits a peak structure,
but the width of the peak is broad and $T_{\rm c}$ becomes
highest for $\beta' \approx -1.2$.
As understood from Fig.~\ref{fig1}, the potential is
the off-center type with a couple of shallow minima
at $q_{\bf i} \approx \pm 2$.
Although $T_{\rm c}$ is not always maximum just
in the range of $-1 < \beta' < -\sqrt{3}/2$,
we conclude that the highest $T_{\rm c}$ is obtained in
the vicinity of the rattling potential with
a wide and flat bottom.

In order to consider further the mechanism of the increase of $T_{\rm c}$,
we estimate the effective electron-phonon coupling constant
$\lambda_{\rm n}$ at $T = T_{\rm c}$.
With the use of the phonon spectral function $\rho_{\rm ph}(\omega)$,
$\lambda_{\rm n}$ is evaluated as
\begin{eqnarray}
  \lambda_{\rm n} = \frac{\lambda \omega_{0} }{2} \int^{\infty}_{-\infty}
  \frac{\rho_{\rm ph}(\omega)}{\omega} d\omega.
\label{lambda}
\end{eqnarray}
For the case of $\beta$=$\gamma$=0 (harmonic phonon),
we easily obtain $\lambda_{\rm n}$=$\lambda$
from the definition of $\lambda_{\rm n}$ and the phonon spectral function.
Note, however, that in general, $\lambda_{\rm n}$ is different from $\lambda$.
Thus, the difference between $\lambda_{\rm n}$ and $\lambda$ is considered
to indicate the effect of anharmonicity on the electron-phonon coupling
constant.

The calculated results for $\lambda_{\rm n}$ are shown in Fig. \ref{fig2}(b).
We immediately observe that in the on-center type potential ($\beta' > -\sqrt{3}/2$),
$\lambda_{\rm n}$ is not so enhanced from the bare value of $\lambda$.
When $\beta'$ enters the value of the region of the rattling-type
potential ($-\sqrt{3}/2  \geq \beta' \geq -1 $),
$\lambda_{\rm n}$ is rapidly increased with the decrease of $\beta'$.
When we further decrease the values of $\beta'$,
$\lambda_{\rm n}$ continues to increase
even in the off-center type potential ($\beta' < -1$).
Here we should pay due attention to the validity of the present adiabatic
approximation in the off-center type potential, but
this point will be commented later.
At the present stage, we suggest that the enhancement of $T_{\rm c}$ is
related to the strong coupling state due to anharmonicity.

Concerning the formation of the peak in $T_{\rm c}$,
we provide a qualitative comment from the strong-coupling effect.
In the BCS theory, it is well known that $T_{\rm c}$ is expressed as
$T_{\rm c}=1.13\omega_0 e^{-1/\lambda}$.
When we consider the renormalization effect,
the formula of $T_{\rm c}$ is changed as
$T_{\rm c}=1.13\omega_0 e^{-Z/\lambda_{\rm n}}$,
where $Z$ is the mass enhancement factor.
Namely, $1/Z$ indicates the renormalization effect.
Note that when $\lambda_{\rm n}$ is moderately large,
$Z$ is roughly given by $1+\lambda_{\rm n}$,
leading to the simplest version of
the McMillan formula.\cite{McMillan}
When $\lambda_{\rm n}$ is small, the renormalization effect
is not so significant and the increase of $\lambda_{\rm n}$
contributes to the enhancement of $T_{\rm c}$.
On the other hand, when $\lambda_{\rm n}$ becomes much larger than unity,
the renormalization effect should be dominant and it rather
suppresses the value of $T_{\rm c}$.
Thus, the peak is formed in $T_{\rm c}$ as a function of $\beta'$.

\begin{figure}
\begin{center}
\includegraphics[width=80mm]{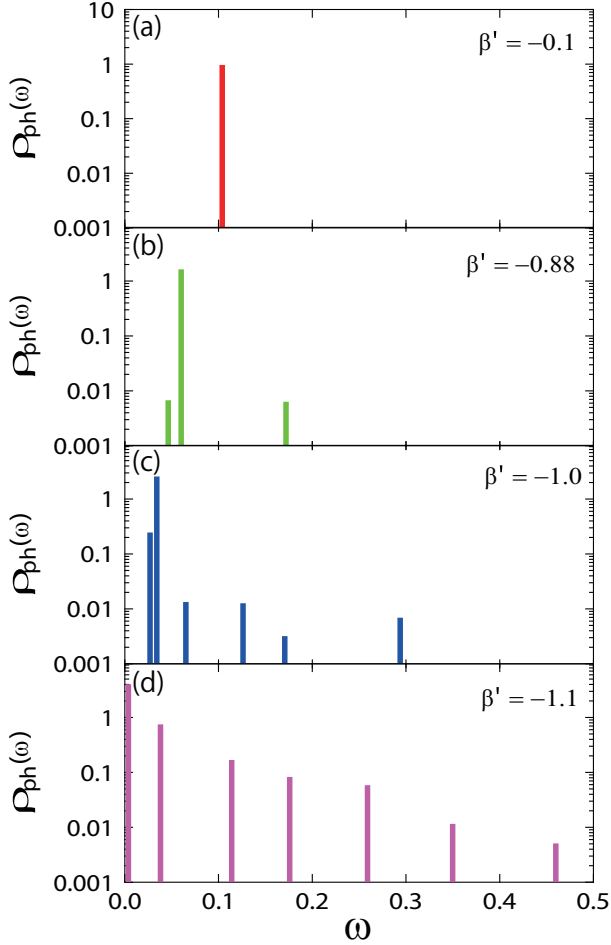}
\end{center}
\caption{(Color online)
Phonon spectral function $\rho_{\rm ph}(\omega)$  vs. energy $\omega$
for (a) $\beta'=-0.1$, (b) $\beta' = 0.88$, (c) $\beta' = -1.0$,
and (d) $\beta' = -1.1$.
Other parameters are set as $\lambda=0.5$, $\omega_{0}=0.1$, and $\gamma=0.001$.
Note that we calculate $\rho_{\rm ph}(\omega)$ at $T = T_{\rm c}$.
}
\label{fig3}
\end{figure}

\subsection{Phonon softening due to anharmonicity}

In the previous subsection, it has been clarified
that the enhancement of $T_{\rm c}$ is triggered
by the strong-coupling tendency due to anharmonicity.
Next we attempt to understand the origin of such tendency.
For the purpose, let us examine the change of
the phonon spectral function,
when we change the value of $\beta'$.
In Figs. \ref{fig3}, we depict $\rho_{\rm ph}(\omega)$
for $\beta'$=$-0.1$, $-0.88$, $-1.0$, and $-1.1$
at $T =T_{\rm c}$ with $\lambda$=0.5, $\omega$=0.1, and $\gamma$=0.001.
The panel (a) for $\beta' = -0.1$ indicates the results
for the on-center type potential.
We observe that
the phonon spectral function is quite similar to that
of the harmonic phonons with a finite weight
at $\omega$=$\omega_0$.
Then, $\lambda_{\rm n}$ for this case is almost equal to
$\lambda$, as observed in Fig.~\ref{fig2}(b).

When we decrease the value of $\beta'$,
the phonon state with finite weights are gradually shifted to
zero, as observed in Figs.~3(b) and 3(c),
which are the results for the rattling-type potential.
It is easy to understand that $\lambda_{\rm n}$ becomes large,
since low-energy phonon states with finite weights
contribute to $\lambda_{\rm n}$
through $\rho_{\rm ph}(\omega)/\omega$ in eq.~(\ref{lambda}).
In Fig.~\ref{fig3}(d) for the case of $\beta'=-1.1$, the largest weight of
$\rho_{\rm ph}(\omega)$ shifts to the lower energy and
the moderate weights appear in higher $\omega$.
Thus, $\lambda_{\rm n}$ becomes large in the parameter region
of the off-center type potential.
In short, the increase of the anharmonicity leads to phonon
softening. Namely, the weight of $\rho_{\rm ph}(\omega)$ is
shifted to lower energy and thus, $\lambda_{\rm n}$ is enhanced
in the anharmonic potential.
Note that our conclusion is consistent with the previous
result.\cite{Mitsumoto}

\begin{figure}
\begin{center}
\includegraphics[width =
 85mm,angle = 0]{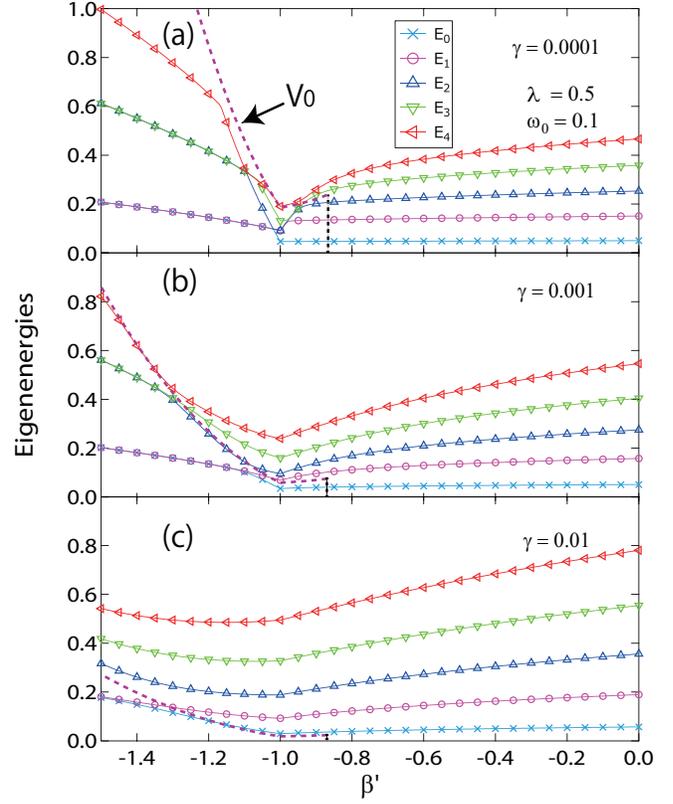}
\end{center}
\caption{(Color online)
Five lowest eigenenergies $E_{0} \sim E_{4}$ (solid curves
with symbols) and $V_{0}$ vs. $\beta'$ (broken curve)
for (a) $\gamma$=$0.01$, (b) $0.001$, and (c) $0.0001$.
Note that we do not show all eigenenergies
in these figures. 
We set $\lambda = 0.5$ and $\omega_{0} = 0.1$.
Since $V_{0}$ dose not exist for $\beta' > -\sqrt{3}/2$
due to its definition,
$V_{0}$ is depicted in the region of $\beta' \leq -\sqrt{3}/2$.
}
\label{fig4}
\end{figure}

Next we discuss the phonon excitation in order to understand
the relation between $T_{\rm c}$ and anharmonic oscillation.
In Figs.~\ref{fig4}(a)-\ref{fig4}(c), we depict eigenenergies $E_0 \sim E_4$ of
$H_{\rm ph}$ and $V_{0}$ for $\gamma$=$0.0001$, $0.001$, and $0.01$,
respectively. 
First let us consider overall tendency of the change of eigenenergy due
to the anharmonicity found in Figs.~\ref{fig4}(a)-\ref{fig4}(c). 
In the case of the on-center type potential for $\beta' > -\sqrt{3}/2$,
all eigenenergies are gradually decreased with the decrease of $\beta'$.
For $-1.0 < \beta' < -\sqrt{3}/2$, in which we observe the rattling-type
potential, the width between eigenenergy levels becomes small.
This is the origin of the shift of phonon spectral function,
as observed in Fig.~3.
For $\beta' < -1.0$, the eigenenergies become doubly degenerate,
which is originating from the left and right deep minima
in the off-center type potential.
The width between degenerate energy levels are rather large,
when we further decrease $\beta'$, but the degenerate phonon states
lead to the strong coupling state in the present adiabatic approximation.
This point will be commented later.

Intuitively, the phonon softening due to anharmonicity is
understood from the energy of oscillator confined in a potential.
If the potential size becomes wide, the momentum of the oscillator
becomes small and the kinetic energy is suppressed
due to the well-known uncertainty principle.
In the present case, the size of the potential bottom is
effectively enlarged due to the anharmonicity and thus,
the typical phonon energy is suppressed due to the quantum effect.
We believe that it is the main effect of anharmonicity on
the electron-phonon coupling state.

Next we turn our attention to the change of $V_0$,
since $V_0$ competes with the zero-point energy $E_0$.
Note again that $V_0$ is defined as the difference between
the potential values at local maxima and minima.
From the competition between $V_0$ and $E_0$,
we can discuss the quantum effect of potential shape,
when there exists structure in the bottom of the potential.
For $\gamma$=$0.0001$ in Fig. \ref{fig4}(a),
since $V_{0}$ is always larger than $E_{0}$ and $E_{1}$,
the guest ion oscillates in the narrow width of
the potential bottom, which is almost the same as that of
the harmonic potential in the low-temperature region.
On the other hand, for $\beta' < -1.0$,
since the phonon energies are degenerate,
the oscillation state is drastically changed
from the small oscillation at the potential minimum to
the quantum tunneling or thermal activation between
two potential minima.
Thus, the behavior of $T_{\rm c}$ is rapidly changed
at $\beta'$=$-1.0$ in the present calculations.
For $\gamma$=$0.001$ in Fig. \ref{fig4}(b),
$V_{0}$ is close to $E_{0}$ at $\beta' \simeq -1.0$,
corresponding to the peak observed in the curve for
$T_{\rm c}$ in Fig.~2(a).
For $\gamma$=$0.01$, $V_{0}$ crosses $E_{0}$
at $\beta' \simeq -1.2$, which also corresponds to the peak
position in $T_{\rm c}$.

From these results, we consider that the behavior of $T_{\rm c}$
is closely related to $E_{0}$ and $V_{0}$.
For $\gamma$=$0.001$ and $0.01$, it is found that
$T_{\rm c}$ takes the maximum value at the parameters
corresponding to $E_{0} \simeq V_{0}$.
In this case, the oscillation amplitude of guest ion is
rapidly increased, since the ion can feel the potential structure
in the energy region less than the zero-point energy.
Namely, the condition of $E_{0} \simeq V_{0}$ just indicates
the rattling oscillation of the guest ion.
Then, $T_{\rm c}$ is considered to take
the maximum value in the rattling type potential.

Hattori and Tsunetsugu have discussed the variation of $T_{\rm c}$
by using three-dimensional anharmonic phonons in the tetrahedral
symmetry.\cite{Hattori}
They have found a crossover in the energy spectrum to the quantum
tunneling regime and $T_{\rm c}$ becomes large at the crossover point.
We consider that their condition of the crossover point corresponds
to $E_{0} \simeq V_{0}$.
Thus, our result is consistent with the previous study.

\subsection{Zero-point energy and anharmonicity}

Now we focus on the relation between the zero-point energy $E_{0}$ and
the potential height $V_0$.
When the local maximum dose not exist for $\beta' = -\sqrt{3}/2$,
$V_{0}$ is defined by the potential height at the saddle point.
For $-\sqrt{3}/2 < \beta'$, $V_{0}$ cannot be defined.
It should be noted that $E_{0}$ is the lowest energy for
the oscillation of the guest ion.

For the rattling-type potential, the potential has minimum
at $q_{\bf i}=0$.
At low temperatures, for $E_{0} < V_{0}$, the guest ion oscillates in
the narrow width of the potential bottom, which is almost the same for
that of the harmonic potential.
On the other hand, for $E_{0} > V_{0}$, the guest ion can oscillate
with large amplitude, since the potential is wide for the
energy larger than $V_{0}$.
Also in the off-center type potential, for $E_{0} > V_{0}$, the guest
ion can oscillate with large amplitude,
since the structure at the bottom is masked by the zero-point energy.
However, for $E_{0} < V_{0}$, the guest ion oscillates in one of
potential minima and sometimes moves to other
potential minima due to quantum tunneling effect.
Thus, it is considered that the oscillation of the guest
ion in the anharmonic potential is quite different
between two cases of $E_{0} > V_{0}$ and $E_{0} < V_{0}$.

\begin{figure}
\begin{center}
\includegraphics[width = 85mm]{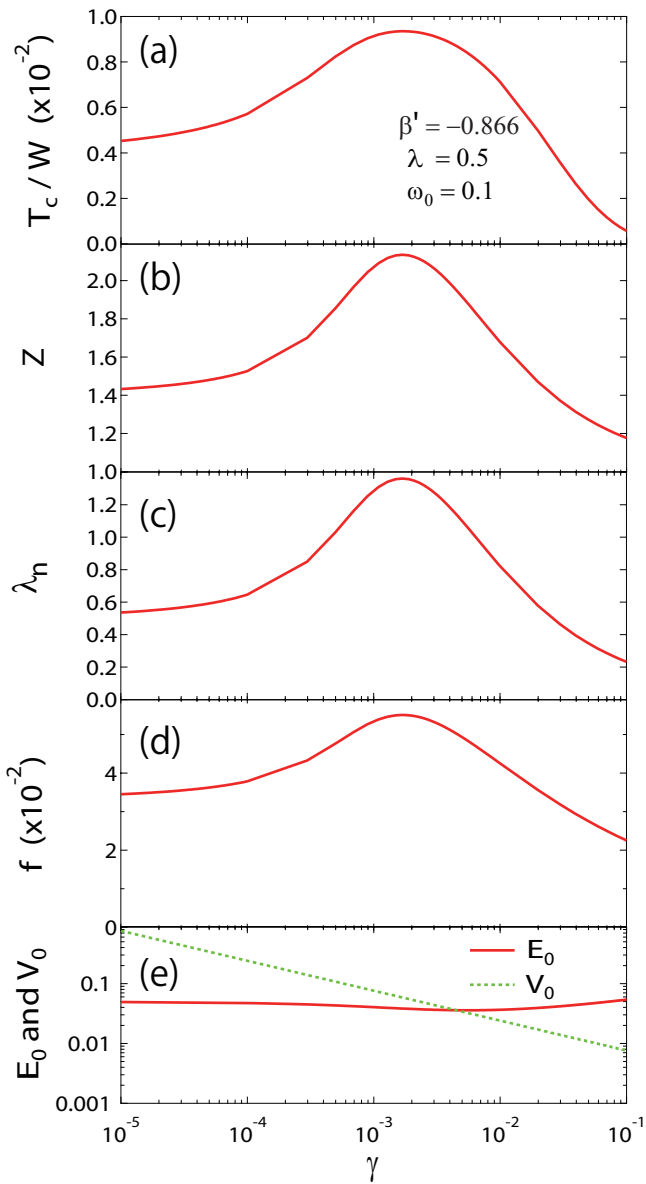}
\end{center}
\caption{(Color online) (a) Superconducting transition temperature
$T_{c}$ vs. $\gamma$.
(b) Mass enhancement factor $Z$ vs. $\gamma$.
(c) Effective electron-phonon coupling constant $\lambda_{\rm n}$
at $T = T_{\rm c}$ vs. $\gamma$.
(d) Non-dimensional Debye-Waller factor $f$ vs. $\gamma$.
(e) The zero-point energy $E_{0}$ and the potential height
at the saddle point $V_{0}$ vs. $\gamma$.
For these panels, we always keep the relation of $\beta' = -\sqrt{3}/2$.
Other parameters are set as $\lambda$=0.5 and $\omega_{0}$=0.1.
}
\label{fig5}
\end{figure}

Here let us discuss $T_{\rm c}$ by focusing on the potential at
$\beta' = -\sqrt{3}/2$, which is depicted by thick curve in Fig.~\ref{fig1}.
Note again that $V_{0}$ is the potential height at the saddle point
in this case and $\beta$ depends on $\gamma$ through the relation of
$\beta = -\sqrt{3\gamma}/2$.
In Figs.~\ref{fig5}(a)-\ref{fig5}(c), we depict $\gamma$ dependence of
$T_{\rm c}$, $Z$, and $\lambda_{\rm n}$, respectively.
When we increase the value of $\gamma$ from $10^{-5}$,
all those values are increased, but
they turn to be decreased to form peaks at a certain value of $\gamma$,
which is defined as $\gamma^*$.

In order to elucidate the relation with the amplitude of the
guest ion, we evaluate the Debye-Waller factor which represents the
intensity of thermal motion.
In the non-dimensional form, it is given by
\begin{eqnarray}
 f = \frac{F_{\rm DW}}{|\mib{G}|^{2} \ell^{2}} = \frac{\langle Q_{\bf i}^{2}
 \rangle}{3 \ell^{2}},
\end{eqnarray}
where $F_{\rm DW}$ denotes the Debye-Waller factor, $\mib{G}$ denotes
the reciprocal lattice vector, $\ell$ is the typical length scale
defined by $\ell$=$g/\omega_{0}^{2}$,
and $\langle \cdots \rangle$ denotes the
operation to take thermal average.
In Fig. \ref{fig5}(d), we show the result for $f$.
We find that $f$ exhibits the peak structure which is quite similar
to those in Figs.~\ref{fig5}(a)-\ref{fig5}(c).
Namely, $f$ is closely related to the behavior of $T_{\rm c}$,
$Z$, and $\lambda_{\rm n}$.

In Fig. \ref{fig5}(e), we depict two curves of $E_{0}$ and $V_{0}$.
At $\gamma = \gamma^{*}$, they cross with each other, so that the guest
ion oscillates with large amplitude in the case of $E_{0} \simeq V_{0}$.
Thus, the large amplitude of the guest ion leads to the
strong coupling state and the enhancement of $T_{\rm c}$.
In addition, it is found that this oscillation causes
the heavy electron state.
This is consistent with our previous result.\cite{oshiba}

\subsection{Debye-Waller factor and $T_{\rm c}$}

Let us now try to understand the tendency of the variation of
$T_{\rm c}$ experimentally observed in $\beta$-pyrochlore oxides
AOs$_{2}$O$_{6}$ (A = K, Rb, Cs)
by focusing on the Debye-Waller
factor at the room temperature and the anharmonicity.

First we briefly summarize the experimental facts
for $\beta$-pyrochlore oxides.
The isotropic atomic displacement parameter $U_{\rm iso}$
at the room temperature which corresponds to the Debye-Waller factor
in our calculation increases as
$2.5 \times 10^{-4}$${\rm nm}^{2}$,
$3.41 \times 10^{-4}$${\rm nm}^{2}$,
and $7.4 \times 10^{-4}$${\rm nm}^{2}$
in the order of decreasing ionic radius, i.e., Cs, Rb, and K.\cite{Hiroi2}
On the other hand, $T_{\rm c}$ and $\lambda_{\rm n}$
for $\beta$-pyrochlore oxides are summarized as
$T_{\rm c}$=3.25 K and $\lambda_{\rm n}$=2.76 for Cs,
$T_{\rm c}$=6.4 K and $\lambda_{\rm n}$=3.38 for Rb,
and $T_{\rm c}$=9.6 K and $\lambda_{\rm n}$=6.3 for K.
Note that the experimental value of $\lambda_{\rm n}$ is obtained
by $1+\lambda_{\rm n}$=$\gamma_{\rm exp}/\gamma_{\rm b}$,
where $\gamma_{\rm exp}$ is the experimental Sommerfeld constant and
$\gamma_{\rm b}$ is evaluated by the band-structure calculations.
Moreover, Kune\v{s} {\it et al.} have clarified that the anharmonicity
of the potential which A ion feels increases in the order of
Cs, Rb, and K.\cite{Kunes}
Thus, in $\beta$-pyrochlore oxides, $T_{\rm c}$ and $\lambda_{\rm n}$ are
closely related to the anharmonicity.

In Figs.~6(a) and 6(b),
we show the calculated results of $T_{\rm c}$ vs. $f$ and
$\lambda_{\rm n}$ vs. $f$, respectively.
Here we change the anharmonicity parameters ($\beta$, $\gamma$)
on the straight line which connects the points of
($\beta$, $\gamma$)=($-0.00392$, $0.00154$) and ($-0.00253$, $0.00451$).
Namely, since the non-dimensional anharmonicity parameter $\beta'$
is changed from $-0.35$ to $-1.0$, the potential shape is changed
from on-center type to rattling one.
It is understood that when $f$ is increased, both $T_{\rm c}$ and
$\lambda_{\rm n}$ are increased.
This tendency agrees well with the experimental situation.
Note that it is easy to find other parameter regions
to depict the figure similar to Fig.~6.
It is concluded that in the rattling potential with
large amplitude of the guest ion, $T_{\rm c}$ and the mass enhancement
is increased with the increase of the Debye-Waller factors,
as observed in $\beta$-pyrochlore oxides.

\begin{figure}
\begin{center}
\includegraphics[width = 85mm]{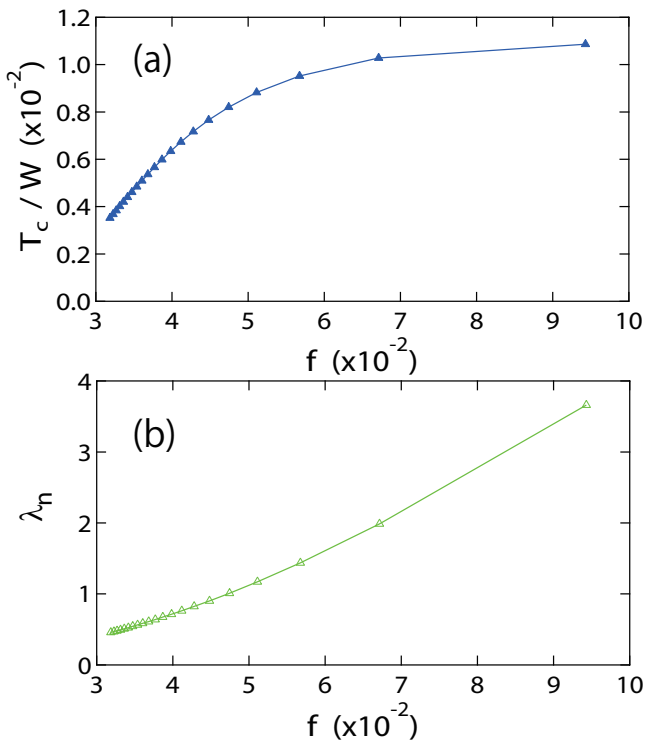}
\end{center}
\caption{(Color online) (a) Superconducting transition temperature $T_{\rm c}$
vs. $f$ and (b) $\lambda_{\rm n}$ vs. $f$.
Concerning the parameters to depict these curves, see the main text.
}
\label{fig6}
\end{figure}

\section{Discussion and Summary}

In this paper, we have analyzed the anharmonic Holstein model
which describes the coupling of conduction electrons with
local ion oscillations in the potential with
fourth- and sixth-order anharmonic terms,
in order to understand the enhancement of $T_{\rm c}$
due to anharmonicity.
First we have investigated $T_{\rm c}$ and the effective electron-phonon
coupling constant $\lambda_{\rm n}$.
Then, we have found that $T_{\rm c}$ is enhanced by the strong coupling
tendency due to anharmonicity.
When the guest ion oscillates with large amplitude,
the phonon softening occurs.
The phonon spectral weights are shifted towards lower energy
and the strong coupling situation is realized even if we consider
originally the weak-coupling case.
This strong-coupling tendency due to anharmonicity is considered
to be a route to arrive at high $T_{\rm c}$.
We believe that this type of strong-coupling superconductivity
is realized in $\beta$-pyrochlore oxides.
The relation between $T_{\rm c}$ vs. $f$ as well as $\lambda_{\rm n}$
vs. $f$ seems to support our present proposal.

In the present calculation, we have always used the same phonon energy $\omega_0$,
but in actuality, $\omega_0$ is different from material to material.
The energy of A ion has been reported in the specific heat measurements,\cite{Hiroi4}
the Raman scattering experiment \cite{Hasegawa}
and the neutron scattering experiment.\cite{Sasai,Mutka}
Nagao {\it et al.} have reported that
the energies of the oscillator are 75.1K for A=Cs,
66.4K for A=Rb, and 22K and 61K for A=K
in $\beta$-pyrochlore oxides AOs$_{2}$O$_{6}$.\cite{Hiroi4}
If the oscillation energy for the case of A=K is taken as 61K,
roughly speaking, it is validated to use the same $\omega_0$
for three $\beta$-pyrochlore oxides.
However, if we pay our serious attention to the phonon energy of 22K,
we should consider a situation in which the oscillators for A=Cs and Rb
have same energy, while the energy for A=K is smaller than it.
If we attempt to pursue quantitative agreement with the experimental facts,
it may be necessary to consider the difference among materials.
However, the research along such a direction should be done with
the use of more realistic model.
It is one of future problems.

In the present adiabatic calculations, $\lambda_{\rm n}$ is monotonically
increased with the decrease of $\beta'$, as shown in Fig.~2(b).
We consider that the calculations are valid for the on-center type,
the rattling-type, and the off-center type potential with shallow minima.
However, when the value of $\lambda_{\rm n}$ is enhanced
in the off-center type potential with deep minima
at non-zero $q_{\bf i}$'s,
another effect should be significant.
Namely, for large $\lambda_{\rm n}$,
the effective bandwidth given by $W/Z$ becomes comparable with
the phonon energy and the adiabatic condition will be violated.
Thus, in the strong-coupling region, it is essential to
include the non-adiabatic effect through electron-phonon
vertex corrections.
When we include such non-adiabatic effect,
the degeneracy in phonon energy originating from the right and left positions
of the off-center type potential will be lifted by the electron-phonon coupling.
In the present calculation, unfortunately, the vertex corrections are not
included and the results in the region of $|\beta'| \gg 1$ with negative $\beta'$
should be reconsidered with the inclusion of the vertex corrections.
This is another future task.

In summary, we have found that the large amplitude anharmonic
oscillation induces the enhancement of $T_{\rm c}$ and
the heavy electron state through the strong-coupling tendency
controlled by the anharmonicity.
We have shown that when the Debye-Waller factor is increased,
both $T_{\rm c}$ and $\lambda_{\rm n}$ are increased.
This tendency agrees qualitatively well with the experimental facts
in $\beta$-pyrochlore oxides.

\acknowledgment

The authors thank T. Fuse for discussions on rattling phenomena.
This work has been supported by a Grant-in-Aid for Scientific Research
on Innovative Areas ``Heavy Electrons''
(No. 20102008) of The Ministry of Education, Culture, Sports,
Science, and Technology, Japan.


\end{document}